\documentclass[a4paper,11pt]{article}
\pdfoutput=1 

\usepackage{jheppub} 

\usepackage[T1]{fontenc} 
\usepackage{amssymb} 

\newcommand{\beq}{\begin{equation}}
\newcommand{\eeq}{\end{equation}}
\newcommand{\bea}{\begin{eqnarray}}
\newcommand{\eea}{\end{eqnarray}}
\newcommand{\ql}{\ensuremath{q_{\,l}}}
\newcommand{\sig}{{\cal S}}

\newcommand{\Dz}{{\cal D}_0}

\newcommand{\Ncb}{{\cal N}_{cb}}
\newcommand{\Ncc}{{\cal N}_{cc}}
\newcommand{\Nbb}{{\cal N}_{bb}}
\newcommand{\Nlb}{{\cal N}_{lb}}
\newcommand{\Nlc}{{\cal N}_{lc}}
\newcommand{\bg}{{\cal B}}
\newcommand{\Ba}{\ensuremath{{\cal B}_1}}
\newcommand{\Bb}{\ensuremath{{\cal B}_2}}
\newcommand{\BF}{{\cal BF}}

\newcommand{\Rlc}{{\cal R}_{lc}}
\newcommand{\Rcb}{{\cal R}_{cb}}
\newcommand{\Rcc}{{\cal R}_{cc}}
\newcommand{\Rlb}{{\cal R}_{lb}}

\newcommand{\Vcb}{\ensuremath{|V_{cb}|}}
\newcommand{\Vcbsq}{\ensuremath{|V_{cb}|^2}}

\newcommand{\lum}{{\cal L}}
\newcommand{\epspre}{\ensuremath{\epsilon_0}}
\newcommand{\eflav}{\ensuremath{\epsilon_B^2}}
\newcommand{\epsflav}{\ensuremath{\epsilon_B^2}}

\newcommand{\epsB}{\ensuremath{\epsilon_{B}}}
\newcommand{\epsl}{\ensuremath{\epsilon_{l}}}
\newcommand{\epsb}{\ensuremath{\epsilon_{b}}}
\newcommand{\epsc}{\ensuremath{\epsilon_{c}}}

\newcommand{\fia}{\ensuremath{f_{i\alpha}}}

\newcommand{\fcb}{\ensuremath{f_{cb}}}
\newcommand{\fll}{\ensuremath{f_{ll}}}
\newcommand{\fcc}{\ensuremath{f_{cc}}}
\newcommand{\fbb}{\ensuremath{f_{bb}}}
\newcommand{\fbc}{\ensuremath{f_{bc}}}

\newcommand{\flb}{\ensuremath{f_{lb}}}
\newcommand{\flc}{\ensuremath{f_{lc}}}
\newcommand{\fcl}{\ensuremath{f_{cl}}}
\newcommand{\fbl}{\ensuremath{f_{bl}}}
\newcommand{\fcbt}{\ensuremath{\widetilde{f}_{cb}}}
\newcommand{\ttbar}{\ensuremath{t\overline{t}}}
\newcommand{\tbar}{\ensuremath{\overline{t}}}
\newcommand{\bbar}{\ensuremath{\overline{b}}}

\newcommand{\qbar}{\ensuremath{\overline{q}}}
\newcommand{\bi}{\begin{itemize}}
\newcommand{\ei}{\end{itemize}}
\newcommand{\delS}{\delta_{\epsb}}

\newcommand{\delBi}{\delta_{\bg i}}
\newcommand{\delBa}{\delta_{\Ba}}
\newcommand{\delBb}{\delta_{\Bb}}
\newcommand{\sigBeps}{\ensuremath{1.3\times 10^4}}
\newcommand{\sigFac}{\ensuremath{23}}
\renewcommand{\ni}{\noindent}
\newcommand{\wlc}{\ensuremath{W\rightarrow\ql\,c}}
\newcommand{\wcc}{\ensuremath{W\rightarrow "c\,c"}}
\newcommand{\wlb}{\ensuremath{W\rightarrow\ql\,b}}
\newcommand{\wbb}{\ensuremath{W\rightarrow "b\,b"}}
\newcommand{\wll}{\ensuremath{W\rightarrow \ql\,\ql}}
\newcommand{\simlt}  {\raisebox{-.6ex}{$\stackrel{\textstyle <}{\sim}$}}

\title{\boldmath A Method to Determine $|V_{cb}|$ at the Weak Scale in Top Decays at the LHC}

\author{P. F. Harrison}
\author{and V. E. Vladimirov}
\affiliation{Department of Physics, University of Warwick,\\Coventry CV4 7AL, United Kingdom}

\emailAdd{p.f.harrison@warwick.ac.uk}
\emailAdd{vangelis.vladimirov@warwick.ac.uk}

\abstract{Until now, the Cabibbo Kobayashi Maskawa matrix element, $|V_{cb}|$, has always been measured in $B$ decays, i.e.~at an energy scale $q_b\sim \frac{m_b}{2}$, far below the weak scale. We consider here the possibility of measuring it close to the weak scale, at $q_W\sim m_W$, in top decays at the Large Hadron Collider (LHC). Our proposed method would use data from the LHC experiments in hadronic top decays $t\rightarrow bW\rightarrow b\bbar c$, tagged by the semileptonic decay of the associated top. We estimate the uncertainty of such a measurement, as a function of present and potential future experimental jet flavour-tagging performances, and conclude that first measurements using the data collected during 2016 - 2018 could yield a fractional error on \Vcb\ of order 7\% per experiment. We also give projected performances at higher luminosities, representative of LHC Run 3 and HL-LHC.}

\begin{document}
\maketitle
\flushbottom

\section{Motivation}
\label{sec:intro}
The $|V_{cb}|$ element of the Cabibbo Kobayashi Maskawa (CKM) quark mixing matrix is currently measured with an uncertainty of about 2\% \cite{PDG}:
\beq
\Vcb =(42.2\pm0.8)\times 10^{-3}.
\eeq
To date, 
it has always been measured in $B$ decays, i.e.~at an energy scale $q\simeq \frac{m_b}{2}$, far below the weak scale. Here, we present a study exploring the feasibility of measuring $|V_{cb}|$ using data from the LHC at the scale $q\simeq m_W$, more representative of the weak scale. The interest in such a measurement is that the traditional extraction of \Vcb\ at the scale of $B$ decays relies heavily on the operator product expansion, and its sensitivity is significantly affected by theoretical uncertainties \cite{PDG}. In contrast, in dealing with decays of on-shell $W$s, as here, the theoretical situation is likely to be much cleaner and the experimental systematic uncertainties will also be very different. Moreover, the value of \Vcb\ measured at the weak scale could be significantly affected by new physics contributions \cite{DFFGV}.

\section{Overview of the Method}
\label{sec:analysis}
We propose to measure $\Vcbsq $ in the decays of tagged \ttbar\ pairs with one semileptonic top decay (the tag), $\tbar\rightarrow \bbar W^- \rightarrow \bbar \ell^- \overline{\nu}_{\ell}$, and the other a hadronic decay, $t\rightarrow bW\rightarrow b\overline{b}c$ (charge-conjugate decays will be assumed everywhere unless otherwise stated). Thus our signal sample is composed of events with three tagged $b$-jets and a tagged $c$-jet, in addition to a charged lepton and missing transverse momentum. As leptons, $\ell^{\pm}$, we consider only electrons and muons, since the majority of $\tau$s decay hadronically.\footnote{The small fraction of taus which decay to electrons and muons may contribute a small amount to our measured rate.} 

Among the hadronic decays of a single top quark, exactly half have a $c$-quark in the final state (up to negligible phase-space factors), and among those, the probability to also have a $\overline{b}$-quark is, by definition, $\Vcbsq \simeq (0.042)^2\simeq 1.8\times 10^{-3}$. Thus we have:
\beq
\Vcbsq=\frac{\Gamma(t\overline{t} \rightarrow b\overline{b}c, \ell^-\overline{\nu} \bbar)}{\Gamma(t\overline{t} \rightarrow  b\overline{q}c, \ell^-\overline{\nu} \bbar)}=
\frac{{\cal BF}(t\overline{t} \rightarrow b\overline{b}c, \ell^-\overline{\nu} \bbar)}{{\cal BF}(t\overline{t} \rightarrow b\overline{q}c, \ell^-\overline{\nu} \bbar)}
,\quad {\rm with}~q=d,s,~{\rm or}~b.
\label{eq:ratioDef}
\eeq
Using the ratio of Eq.~(\ref{eq:ratioDef}), otherwise leading experimental uncertainties in most of the tagging efficiencies are cancelled. 
We have considered other approaches, where different numbers of $b$-tags or the $c$-tag are not explicitly required, and have concluded that the approach described above has the best sensitivity. Our approach requires all our hadronic $W$ candidates to be fully-resolved into two jets, and an appropriate reduction in efficiency is included to account for this in our calculations below.

In order to select \ttbar\ semi-leptonic events, we require both numerator and denominator events to satisfy kinematic and flavour identification criteria as follows. Using missing $p_T$ and requiring the neutrino and charged lepton to reconstruct a $W$ mass, we estimate the neutrino momentum. We require the resulting leptonic $W$ and a $b$-tagged jet (denoted here a ``bachelor'' $b$) to be consistent with the top mass. On the hadronic side of the event, we require that two jets, one of them $c$-tagged, are consistent with a $W$, and that combined with another (bachelor) $b$-tagged jet they are consistent with a top. These requirements suppress non-$t\tbar$ backgrounds substantially. For the numerator, we require additionally that one of the $W$-daughter jets is explicitly flavour-tagged as a $b$-jet (the ``signal'' $b$).

\section{Sample Magnitude Estimates}
\label{sec:ctagged}
Both the ATLAS and CMS experiments trigger top-pair events with good efficiency and have published total and differential \ttbar\ production cross-section measurements at 13 TeV collision energy \cite{ATLAStotal13, CMStotal13, ATLASdifferential, CMSdifferential}. In particular, averaging the measurements, the total top-pair production cross-section at 13 TeV is $855 \pm 25$ pb, meaning that the total Run 2 (2016-2018) data set at each LHC experiment should correspond to $\sim 10^7$ reconstructed \ttbar\ events. 

In order to estimate the denominator and numerator sample sizes, we list in table \ref{table:signalMag}, the quantities on which they depend, together with the variable names we use, their values and their origins. In much of what follows, we leave the integrated luminosity as a free parameter, but note that the total from Run 2 is ${\cal L}\simeq 140$~fb$^{-1}$ per experiment.

Apart from our jet flavour-tagging requirements, the proposed analysis follows quite closely published ATLAS and CMS $\ttbar$ total cross-section analyses in the same lepton-plus-jets mode. Thus, the scale of our pre-selection efficiency is likely to be set by that in those analyses. By pre-selection efficiency, we mean the experimental efficiency to accept, trigger, reconstruct and select the events which enter the denominator in eq.~(\ref{eq:ratioDef}), and include the lepton tagging efficiency, but exclude the jet flavour-tagging. Published ATLAS and CMS \ttbar\ measurements obtained top-pair pre-selection efficiencies in the region of 17\% for ATLAS at 8 TeV~\cite{ATLAStotal8}, and 13\% for CMS at 13 TeV~\cite{CMStotal13} (the selection cuts were different in the two cases; neither includes any branching fractions). For the purposes of estimating the sample sizes for this measurement, we set it conservatively to $\epspre=10\%$ here. We have verified the kinematic part of this with a stand-alone fast event simulation. The jet flavour-tagging efficiencies \epsB, \epsc\ and \epsb, are left as free parameters in this study, since they need to be optimised in the full analysis for the sensitivity to \Vcbsq. However, we later give representative values for them, in order to estimate the sensitivity which might be achieved.

\begin{table}[t]
\centering
\begin{tabular}{ l  c  c  c  }
\hline
  Quantity & Symbol & Origin of Value & Approx. Value\\[0.8ex]
\hline
$t\overline{t}$ Cross-section & $\sigma_{t\overline{t}}$ &  \cite{ATLAStotal13}\cite{CMStotal13} & $8.5\times 10^5$ fb \\[0.8ex]
Integrated Luminosity & $\cal{L}$ & Time-variable & 140\,fb$^{-1}$ (Run 2) \\[0.8ex]
$\BF(\ttbar\rightarrow \ell\nu b\bbar c\qbar)$& $\BF$ & $2\times(\frac{2}{9})\times(\frac{2}{3})\times\frac{1}{2}$ &  $0.15$\\[0.8ex]
Lepton Tag Efficiency & $\epsilon_{lep}$ & $-$ & $-$ \\[0.8ex]
Pre-selection Efficiency & $\epspre$ &  \cite{CMStotal13}\cite{ATLAStotal8} Incl.~$\epsilon_{lep}$, not $\epsflav$  & 0.10 \\[0.8ex]
Pre-selected Cross-section & - & $\sigma_{t\overline{t}}\times\BF\times\epsilon_0$ & $\sigBeps$ fb\\[0.4ex]
Bachelor $b$ Flavour Tag Effic. & $\epsB$ & Optimisation & Tunable \\[0.8ex]
Charm Flavour Tag Effic. & $\epsc$ & Optimisation & Tunable \\[0.8ex]
Signal $b$ Tag Efficiency & $\epsilon_b$ & Optimisation &  Tunable \\[0.8ex]
\hline
\end{tabular}
\caption{Input quantities for calculation of sample magnitudes. The two $b$-jet flavour-tagging efficiencies, $\epsB$ and $\epsb$, are kept as distinct quantities, foreseeing the possibility to tune them independently in optimising the signal sensitivity.}
\label{table:signalMag}
\end{table}
The quantities in table \ref{table:signalMag} are combined to estimate two event sample sizes, each as a function of both the integrated luminosity and the various tagging efficiencies (to be optimised): 1) half the number of reconstructed $\ttbar$ events with two $b$-jet tags and a hadronic $W$ decay in the final state (regardless of the flavour-tags of the $W$ daughters):
\beq
{\Dz}=\sigma_{t\overline{t}}\,\,\BF\,\epsilon_0\,{\cal L}\,\eflav\simeq\sigBeps\,\lum \,\,\epsflav.\label{presampleSize}
\eeq
The factor of a half in the definition corresponds to the fraction of hadronic $W$ decay events having a charm quark in the final state. We expect $\Dz$ to be rather well determined experimentally; and 2) the number of reconstructed signal events:
\bea
{\cal S}=\Dz\,\epsc\,\epsb\,\Vcbsq\simeq \sigFac\,\lum \,\,\eflav\,\epsc\,\epsb.
\label{eq:sig}
\eea
From eq.~(\ref{eq:sig}), \Vcbsq\ is simply given by:
\beq
\Vcbsq=\frac{\sig}{\epsc\epsb{\Dz}}.
\label{sigOverBg}
\eeq

To get a rough idea of sample sizes, taking ``typical'' tagging efficiencies for $b$ and $c$ quarks of order 50\%, the $c$-tagged sample size for the whole of Run 2 is in the region of $\epsc\Dz\sim 2.3\times 10^5$ events per experiment, while the number of signal events is a factor $\epsb\Vcbsq$ smaller, i.e.~${\cal S}\sim 200$ events. In the absence of backgrounds, this would correspond to a fractional statistical uncertainty on \Vcb\ (i.e.~half the fractional error on  \Vcbsq) of about 3.5\%. This would be a very interesting first measurement close to  the weak scale.

\section{Backgrounds}
\label{sec:Backgrounds}
Backgrounds in the proposed analysis may be divided into two classes: \ttbar\ backgrounds and non-\ttbar. Among the \ttbar\ backgrounds, the most important is ``leak-through'' from the denominator to the numerator. There are two main ways this can happen: either the light-quark jet can simply be mis-identified as a $b$-jet (we denote this e.g. $\ql\Rightarrow b$), or two mis-identifications may occur simultaneously: $c\Rightarrow b$ and $\ql\Rightarrow c$. We denote these two background samples $\Ba$ and $\Bb$ respectively. This is a dangerous class of background, since all other aspects of the events are identical to the signal, and we know that the sample from which they may arise is about 1000 times (i.e.~$\frac{1}{\epsb\Vcbsq }$) more populous than the signal. We will need to suppress these backgrounds very well, and also to know the amount remaining well, in order to subtract them with minimal uncertainty. 

Another possible, though less serious, background is from events in which the hadronic $W$ decays to two light quarks, with both misidentified, one as a $c$-jet and one as a $b$-jet. However, the probability of this happening is suppressed significantly relative to the previous mechanisms due to smaller mis-tag rates, and it can therefore be considered a sub-dominant background.

In some denominator events, one of the bachelor $b$-jets (the direct $b$-daughters of the $t$ quarks in either of the top decays), when paired with the $c$ quark may potentially satisfy the $W$ reconstruction selection (in addition to, or in preference to, the light quark jet). Such kinematically-ambiguous events are naturally suppressed (a kinematic coincidence is needed), but could still be a potentially dangerous source of background, since the swapped-in $W$ daughter would be a $b$-jet. However in such background events, the swapped-out jet will be a light quark jet treated as a bachelor top daughter so that such events are actively suppressed by our approach of requiring three positive $b$ flavour tags in the event. We have investigated this kind of kinematic ambiguity using a dedicated fast (parameterised) simulation, and found that if there remains residual contamination, it can be rendered negligible by selectively removing events in parts of the top decay Dalitz plot where the kinematics of the three jets are ambiguous, with only a modest additional loss of efficiency.

Other types of background arise from non-\ttbar\ events, and several were studied in existing published \ttbar\ total cross-section analyses by ATLAS~\cite{ATLAStotal8} and CMS~\cite{CMStotal13}. Examples include single-top events and events with a leptonic $W$ decay and jets. While our selection criteria are intended to suppress such backgrounds, the referenced \ttbar\ cross-section measurements indicate that they are at a rather low level ($\simlt\,5\%$) and should be a minor concern here. 

In order to provide a first estimate of the background magnitude, and on the basis of arguments already made, we take it as evident that feed-through from the denominator due to mis-tagging is the most significant source of background for this analysis, and thus we focus in what follows on backgrounds \Ba\ and \Bb. Since our numerator and denominator differ by the flavour of a single quark jet ($b$ compared with $d$ or $s$), the level of these backgrounds is governed entirely by mis-tag probabilities.  We will write as \fia, the (kinematically-integrated) probability to identify a pre-selected jet of flavour $i$ ($=l$ for light-flavour or $c$ or $b$) as one of flavour $\alpha$. The numbers of events in our \Ba\ and \Bb\ samples are then:
\bea
\ql \Rightarrow b:~~~&\Ba&=\Dz\,\epsc\flb,\label{eq:B1}\\
c\Rightarrow b~{\rm and}~\ql\Rightarrow c:~~~&\Bb&=\Dz\,\flc\fcb.\label{eq:B2}
\label{bgDecomp}
\eea
Since these two mechanisms for a denominator event to incorrectly feed into the numerator are essentially independent alternatives, their (relatively small) probabilities can be simply added to estimate the total number of such background events as:
\bea
{\cal B}
&\simeq&\Ba+\Bb\label{bgsplit}\\
&=&\Dz\,\epsc(\flb+\flc \,\fcbt)\label{bg1}\\
&\simeq& \sigBeps\,\lum \,\,\eflav\,\epsc\,(\flb+\flc\,\fcbt),
\label{bg}
\eea
where, for notational convenience, we have defined the re-scaled $c\Rightarrow b$ mis-tag probability:
\beq
\fcbt=\frac{\fcb}{\epsc}\label{fcbt}.
\eeq
From eqs.~(\ref{eq:sig}) and (\ref{bg}), we thus have that:\footnote{We note the necessity to take account of a small additional contribution to the numerator of magnitude $\fbc\fcbt\Vcbsq$ coming from self-cross-feed in genuine signal events via the double mistag $b\Rightarrow c$ and $c\Rightarrow b$. We ignore here this small increase in signal efficiency for the sake of presentational clarity.}
\beq
\frac{\sig}{\bg}\simeq\frac{\epsb\Vcbsq}{(\flb+\flc\,\fcbt)}\label{soverb}
\eeq
confirming that we need the mis-tag rates entering the denominator to be $\simlt~10^{-3}$, as already indicated.

\section{Jet Flavour Tagging}
\label{sec:tagging}
\subsection{Mutually Exclusive Flavour Tagging Outcomes}
\label{subsec:mutExclTagging}
Flavour taggers operate in the three-dimensional space of light flavour (denoted $l$ here), $c$- and $b$-flavours~\cite{ATLASbtag1}. In our approach, it will be necessary to test each candidate jet against each of these hypotheses, and allocate it just a single preferred flavour. We are not, e.g.~interested in whether a jet is consistent with being both $b$- and $c$-flavoured (as many jets can be, in principle). Thus, in order to minimise backgrounds, our proposed analysis must work with mutually exclusive tagging outcomes (among the complete set $\left\{l,c,b\right\}$). Given these provisions, it is an obvious step to set-up a "matrix" of tagging probabilities:
\begin{align}
f=\left ( \begin{matrix} 
	\fll & \flc & \flb \\
	\fcl & \fcc & \fcb \\                    
	\fbl & \fbc & \fbb
\end{matrix} \right )
\equiv\left ( \begin{matrix} 
	\epsl & \flc & \flb \\
	\fcl & \epsc & \fcb \\                    
	\fbl & \fbc & \epsb\label{unitarity0}
\end{matrix} \right ).
\end{align}

The diagonal elements are identified as flavour-tagging efficiencies: $f_{ii}\equiv\epsilon_i$, the probability for a reconstructed jet to be correctly tagged as flavour $i$ (in practice, we use the latter, conventional notation for them). The off-diagonal elements, the mis-tag rates, are identified with the reciprocals of the ``flavour-rejection'' rates \cite{ATLASbtag1}. Above the diagonal, the mis-identification has the sense to increase the mass of the identified flavour.  As already discussed, these are the ``dangerous'' mis-identifications for our extraction of \Vcb.

Since the nine \fia\ are to be interpreted as probabilities, unitarity is respected across each row, leading to three constraints:
\beq
\sum_{\alpha} \fia = 1.
\label{unitarity}
\eeq
They simply state that a reconstructed jet born as flavour $i$ must be tagged as exactly one of the three possible outcomes (there is no analogous unitarity constraint down the columns). Thus, only six of our flavour-tag probabilities are independent. For the rest of this paper, we choose to work only with the efficiencies \epsc, and \epsb, and the mis-tag probabilities \flc, \flb, \fcb\ and \fbc, the remaining three being determined by the constraints, eq.~(\ref{unitarity}).

We remark that the tagging formalism discussed above may be implemented as a function of jet-kinematic variables, or alternatively, just adopted in some specific sample(s) integrated/averaged over their kinematics. For this measurement, we propose the kinematically-integrated approach, as motivated by the discussion of section 6.

\subsection{Experiments' Tagging Performances}
\label{subsec:tagPerf}
As discussed at the end of section 4, the proposed method of measuring \Vcb\ requires very tight flavour-tagging, with, e.g.~light quark mis-identification in the $b$-tagger, $\flb \lesssim 10^{-3}$. ATLAS \cite{ATLASbtag1} and CMS \cite{CMSbtag1} have published particular flavour-tagging ``working points'', namely sets of fixed flavour-tagging criteria which have been optimised and characterised for existing experimental analyses (which may have different requirements from our presently-proposed measurement). Both experiments have also provided continuously-varying tagging-performance characteristics (so-called ROC curves), determined from simulated datasets and calibrated on real data. These quantify the performance of the existing tagging algorithms on a continuum of possible working-points, which may be optimised for this or other analyses. Both experiments have also provided estimates of the uncertainties on their tagging-performance values.
\begin{table}[t]
\centering
\begin{tabular}{ c  c c  c c }
\hline
\multicolumn{5}{c}{Beauty Tagging} \\[0.8ex]
\hline
  Quantity &    ATLAS     		&  ATLAS		   &      CMS   		&   CMS Rel.    	\\[0.8ex]
  	        &   & Rel.~Uncert.  & \cite{CMSbtag1}  &  Uncert. \cite{CMSbtag1}   \\[0.8ex]
\hline
$\epsb$ & 0.55 \cite{ATLASbtag1} & 2--9\%, table 6 of \cite{ATLASbtag2} & 0.45  & 2\%\\[0.8ex]
$\fcb$ & $3\times 10^{-2}$ \cite{ATLASbtag1, ATLASctob} & 17\% \cite{ATLASctob} & $2\times 10^{-2}$ &  30\% \\[0.8ex]
$\flb$ & $6\times 10^{-4}$ \cite{ATLASbtag1, ATLASltob} & 15--30\% \cite{ATLASltob} &  $1\times10^{-3}$  & 10\% \\[0.8ex]
\hline
\multicolumn{5}{c}{Charm Tagging}\\[0.8ex]
\hline
  Quantity &    ATLAS     &   ATLAS 			& CMS                     &   CMS Rel. \\[0.8ex]
  	        &  from fig.~1 of \cite{ATLASctag1} & Rel. Uncert.   & \cite{CMSbtag1}   &  Uncert.~\cite{CMSbtag1}\\[0.8ex]
\hline
 $\epsc$ & 0.25 & Unavailable & 0.31  & 4\% \\[0.8ex]
   \fbc & $0.25$  & Unavailable & $0.25$ & 5\%  \\[0.8ex]
$\flc$ & $3\times 10^{-3}$ & Unavailable & $7\times 10^{-2}$ &   10\%  \\[0.8ex]
\hline
\end{tabular}
\caption{Representative jet flavour-tagging efficiencies and mis-tag rates of suggested extra-tight ``working points'' selected from published ATLAS and CMS tagging performance ROC curves. The ATLAS $b$-tag values are taken from fig.~10 of ref.~\cite{ATLASbtag1}; the values of \fcb\ and \flb\  have been modified according to the ``scale factors'' given in (the last bin of) fig.~9 of \cite{ATLASctob} and from fig.~13 of \cite{ATLASltob} respectively. The CMS values for \epsb, \fcb\ and \flb\ are taken from figs.~16 and 17 of ref.~\cite{CMSbtag1} while the values for \epsc, \fbc\ and \flc\ are interpolated from fig.~19. The CMS values are adjusted for the measured scale factors of figs.~33 and 53 of ref.~\cite{CMSbtag1}, from which the fractional uncertainties were also extracted. We note that the ATLAS fractional uncertainties (where available) do not necessarily correspond to the particular ``working points'' we suggest, but rather to the closest published values. The wide ranges of some are due to $p_T$ and other dependencies, where averaged values are not available.}
\label{table:taggingEfficiencies}
\end{table}

In table~\ref{table:taggingEfficiencies}, we summarise the experiments' published flavour-tagging performances. They correspond to points we suggest on the published tagging-performance continua, as possible ``working points'' (loosely) optimised for the measurement proposed here. It is possible that in the future, even better tagging performances may be achieved, given future innovations in tagging methodology, and calibration sample-sizes. However, we emphasise that these performances are the current, published state-of-the-art, and our method does not rely on unjustified projections of future tagging performance.

We remark that, as far as we know, tagging performances based on mutually-exclusive tagging outcomes have not yet been published by ATLAS or CMS. Our use of the performances as published, although not strictly justified in the sense of true tagging probabilities, is not expected to result in a significant over-estimate of the eventual performance of a scheme in which the probabilistic interpretation is fully-justified. On the contrary, the proposed implementation of probability-conserving tagging criteria seems likely, if anything, to improve the performance overall of our proposed measurement.

\section{Mis-tag Background Control Samples}
\label{sec:mistagbackground}
We consider here alternative observable final states of the top-tagged $W$ sample. The numbers of events in these otherwise suppressed/forbidden final states are primarily determined by the flavour mis-tagging rates appearing in eqs.~(\ref{bg1}-\ref{soverb}), and thus may be used to measure those mis-tag rates from the data. The significant advantage of these control channels is that they are sampled from a jet kinematic distribution which is essentially identical to that of the signal. Hence, the values so-obtained are the exact (kinematically-integrated) values which we need, in order to calculate the respective backgrounds in our signal sample. This data-driven approach thus minimizes the systematic uncertainty in our knowledge of these backgrounds, effectively replacing systematic with statistical uncertainties.

We first consider the sample of preselected \ttbar\ events which additionally have their hadronic $W$-decay daughters flavour-tagged as $c$ and $\ql$. The majority of them derive from the sample of $\Dz$ preselected (ie.~pre-tagged) \hbox{\wlc} decays. Their probability to have the $c$-jet correctly tagged is simply $\epsc$ while that for the light quark is $\epsilon_l=(1-\flc-\flb)$ (see eq.~(\ref{unitarity})), quite close to unity. A small contribution also comes from the double mis-tag $c\Rightarrow \ql$ and $\ql\Rightarrow c$ with probability $\fcl\flc\simeq\flc(1-\epsc-\fcb)$. A significant contribution is also made by (actual) $W\rightarrow \ql\ql$ decays with mis-identification $\ql\Rightarrow c$, occurring with probability $\sim 2\flc$. Contributions from the signal mode are suppressed by $\Vcbsq\simeq(0.042)^2$, and are negligible. Thus we obtain a sample size:
\bea
\Nlc&=&\Dz\,[\epsc(1-2\flc-\flb)+\flc(3-\fcb)]\label{denom}
\label{eq:Nlc}
\eea
from which $\epsc$ can be extracted with a negligible error, given knowledge of $\Dz$, $\flc$, $\fcb$ and $\flb$, 
see table \ref{table:backgrounds}.

We next consider the flavour-tagged pseudo $W$-decay final state ``$c\,c$''. This is of course forbidden in the SM, but can be generated in the $c$-tagged \wlc\ sample by the mis-identification $\ql\Rightarrow c$, which happens with mean probability $\flc$. It could also be generated (less probably) from $W\rightarrow \ql\ql$ decays by the mis-identification $\ql\Rightarrow c$, happening twice, i.e.~with mean probability $\flc^2$. It can also be generated from our signal channel by the mis-tag $b\Rightarrow c$. Thus, the number of events in this sample is:
\bea
\Ncc&\simeq&\Dz\,\epsc(\flc+\Vcbsq\fbc+\frac{\flc^2}{\epsc}).
\label{eq:Ncc}
\eea
Given a knowledge of $\Dz$, $\epsc$ and $\fbc$, together with the roughly-known value of \Vcbsq, this rate determines $\flc$.

The flavour-tagged \wlb\ decay can occur in the SM due to the non-zero CKM matrix element $V_{ub}$, but is suppressed by a factor $\sim$100 relative to our signal and is therefore negligible (at least in the SM). It can also occur due to the mis-identification of the allowed \wlc\ decay with probability $\fcb$, and in the same $W$-decay via the double mis-identification $c\Rightarrow \ql$ and $\ql\Rightarrow b$ with total probability $\fcl\,\flb$. It can also arise due to mis-identification of the allowed \wll\ decay with probability $\simeq 2\flb$. Thus, the number of events in this sample is:
\bea
\Nlb
&\simeq&\Dz\,[\fcb+\flb(3-\epsc-\fcb)],
\label{eq:Nlb}
\eea
where we used eq.~(\ref{unitarity}) to eliminate $\fcl$. Since the first term is strongly dominant, this rate essentially measures $\fcb$.

The flavour-tagged pseudo final state \wbb\ can be generated from the allowed \wlc\ decay via tagging misidentifications with probability $\flb\,\fcb$ and from \wll\ with probability $\flb^2$. It can also be generated from our signal channel by the mis-tag $c\Rightarrow b$. Thus, the number of events in this sample is:
\bea
\Nbb
&\simeq&\Dz\,[\flb(\fcb+\flb)+\Vcbsq\epsb\fcb].
\label{eq:Nbb}
\eea
The leading term is $\fcb(\epsb\Vcbsq+\flb)$ but is too small to be helpful in determining $\flb$, so we will not consider it further.

The main findings of this section are summarised in table \ref{table:backgrounds}. We give the leading and sub-leading tagging probabilities contributing to the (normalised) numbers of events in eqs.~(\ref{denom}), (\ref{eq:Ncc}) and (\ref{eq:Nlb}), together with their approximate values (taken from the second column of table~\ref{table:taggingEfficiencies}), and the total numbers of events themselves, ${\cal N}_{ij}$, assuming an estimated integrated Run~2 luminosity, $\lum=140$ fb$^{-1}$. The last column gives the statistical error on the determination of the leading contribution thus measured, calculated simply as the reciprocal of the square-root of the ${\cal N}_{ij}$. It is notable how small they are compared with the usual uncertainties quoted for tagging efficiencies and mis-tag rates in e.g.~table~\ref{table:taggingEfficiencies}.
\begin{table}[t]
\centering
\begin{tabular}{ c  c  c  c  c  c  c}
\hline
  Obsvd.   & Leadng.  & Approx. & Sub-leadng & Sub-lead.   &  ${\cal N}_{ij}$ & Rel.~Stat \\[0.8ex]
 Mode   &  Contr.   & Value     & Contr.      & Fraction   &   $\simeq$ & Error \\[0.8ex]
\hline
\wlc\ &   $\epsc$ & $0.25$ & $\flc(3-2\epsc)$  & $0.03$ & $1.4\times 10^5$ & $0.0027$ \\[0.8ex]
\wcc\ &   $\flc$ & $3\times 10^{-3}$ & $\Vcbsq\fbc$  & $0.15$ & $4.1\times 10^2$ & 0.05 \\[0.8ex]
\wlb\ & $\fcb$ & $3\times 10^{-2}$ & $\flb(3-\epsc)$   & $0.03$ &$1.7\times 10^4$ & 0.008 \\[0.8ex]
\hline
\end{tabular}
\caption{Summary of usable calibration channels. The statistical uncertainties in the last column are calculated assuming the Run 2 luminosity.}
\label{table:backgrounds}
\end{table}
There will be small systematic uncertainties in the $\fia$ measured thus, from background contributions to the overall sample normalisation, $\Dz$, which we have not discussed in any detail. However, these backgrounds themselves are at a level of not more than a few percent, based on the small contributions of non \ttbar\ backgrounds in the \ttbar\ cross-section analyses \cite{ATLAStotal8}, and their uncertainties will be significantly smaller still, rendering them negligible in the present context.

\section{Extraction of {\boldmath \Vcbsq}}
\label{sec:extraction}
We sketch an approach using the signal channel and the three usable calibration channels listed in table \ref{table:backgrounds} in a single simultaneous extraction of \Vcbsq\ together with the relevant tagging probabilities. We start with the numbers of measured events in the four event classes. We have from eqs.~(\ref{eq:sig}), (\ref{bg1}), (\ref{denom}), (\ref{eq:Ncc}) and (\ref{eq:Nlb}):
\bea
\Ncb=\sig+\bg&\simeq&\Dz\,\epsc(\epsb\,\Vcbsq+\flb+\flc\fcbt)\label{Ncb}\\
\Nlc&\simeq&\Dz[\epsc(1-2\flc)+3\flc] \label{Nlc}\\
\Ncc&\simeq&\Dz\,\epsc(\flc+\fbc\Vcbsq+\frac{\flc^2}{\epsc}) \label{Ncc}\\
\Nlb&\simeq&\Dz\,\epsc[\fcbt+\flb(\frac{3}{\epsc}-1)]\label{Nlb}.
\eea
Taking also the quantity $\Dz$, defined in eq.~(\ref{presampleSize}), we define event number ratios
\beq
\Rlc=\frac{\Nlc}{\Dz}
\eeq
and (for the remaining three event classes)
\beq
{\cal R}_{ij}=\frac{{\cal N}_{ij}}{\Dz\epsc}.
\eeq
In terms of them, we find
\bea
\epsc&=&\frac{\Rlc-3\flc}{1-2\flc-\flb}\label{eq:epsFInal}\\
\flc&=&\frac{1}{2}\epsc\left[\sqrt{1+4\frac{\Rcc'}{\epsc}}-1\right]\simeq\Rcc'(1-\frac{\Rcc'}{\epsc})\label{eq:flcFInal}\\
\fcbt&=&\Rlb-\flb(\frac{3}{\epsc}-1)\label{eq:fcbFInal},
\eea
where
\beq
\Rcc'=\Rcc-\fbc\Vcbsq,\label{RccPr}
\eeq
and finally
\beq
\Vcbsq=(\Rcb-\flb-\flc\fcbt)/\epsb.
\label{eq:combined}
\eeq
Taking \flb, \epsb\ and \fbc\ as external input from (other) tagging studies, the four simultaneous equations~(\ref{eq:epsFInal})-(\ref{eq:combined}) determine \epsc, \flc, \fcbt\ and \Vcbsq\ in terms of the four event number ratios. Their values may be extracted to the desired precision (up to statistical uncertainties in the numbers of events), by appropriate numerical or iterative methods. Table \ref{table:backgrounds}, indicates that the sub-leading contribution to $\Rcc'$, Eq.~(\ref{RccPr}), is of relative magnitude 15\%, whereas its statistical error is of order 5\%. Thus, as long as the systematic error on the sub-leading contribution can be controlled at the $\simlt$15\% fractional level, it will be negligible in comparison to the statistical error on \flc. We expect this to be the case. Under these circumstances, \flb\ and \epsb\ are the only sources of systematic uncertainty entering our extraction equations and we proceed on this basis.

\section{Estimated Experimental Uncertainties and Sensitivity to {\boldmath \Vcbsq}}
\label{sec:errorsandsens}
We estimate what sensitivity to \Vcbsq\ may be possible in the future, given the tagging performances summarised in table \ref{table:taggingEfficiencies} and possible future uncertainties in the $b$-tagging performance. We need first to quantify the sources of statistical and systematic uncertainties which contribute to the overall uncertainty in the measured value of \Vcbsq. We first re-interpret eq.~(\ref{eq:combined}) schematically in terms of signal and background samples (c.f.~eq.~(\ref{sigOverBg})):
\beq
\Vcbsq\simeq\frac{(\sig+\bg)-\Ba-\Bb}{\Dz\,\epsc\epsb},
\label{combined3}
\eeq
where we identify the parenthetical term with $\Ncb$, and \Ba, \Bb\ were defined in eqs.~(\ref{eq:B1}-\ref{eq:B2}).

The first contribution to the uncertainty is the ``standard'' statistical contribution of a signal with a significant background source, associated with the parenthetical term in the numerator (evaluated, {\em before} background subtraction):
\beq
\textrm {Statistical~error}\propto\sqrt{\sig+\bg}\sim\sqrt{\Ncb}.
\label{staterr}
\eeq
Additional uncertainties come from the imperfect subtraction of the backgrounds, $\Ba$ and $\Bb$. For convenience, we parameterise their uncertainties as fractional uncertainties in their magnitudes, $\delBa$ and $\delBb$ times their respective magnitudes $\bg_i$.

The statistical uncertainty on the denominator of eq.~(\ref{combined3}) is negligible by comparison with the others while a systematic contribution comes from the uncertainty in the $b$-tagging efficiency $\epsb$. This contributes a fractional uncertainty in the measurement, parameterised here as a variable $\delS$, the fractional uncertainty on $\epsilon_b$. At the time of writing, the magnitudes of the fractional uncertainties $\delBa$ and $\delS$ are not known, since they depend on the uncertainties in tagging and mis-tag probabilities, which have yet to be optimised for the analysis. In what follows, we leave them as free parameters, anticipating that values $<15\%$ will be achievable, see table \ref{table:taggingEfficiencies}.

Since the mis-tag rates contributing to $\bg_2$, eq.~(\ref{bgDecomp}), are measured as discussed in section \ref{sec:extraction}, and we argued there that their statistical errors dominate, we find:
\beq
\delBb=\sqrt{\frac{1}{\Ncc}+\frac{1}{\Nlb}}.
\eeq

We therefore add in quadrature to the statistical uncertainty in eq.~(\ref{staterr}), a further statistical contribution $\delBb\Bb$, together with systematic contributions of the form $\delS \sig$ and $\delBa\Ba$. 
Our total fractional error on \Vcbsq\ therefore takes the form:
\beq
\frac{\Delta\Vcbsq}{\Vcbsq}\sim\frac{\sqrt{\sig}\oplus\sqrt{\bg}\oplus\delS \sig\oplus\sum_i \delBi \bg_i}{\sig},
\label{totErr}
\eeq
where all terms including those under the summation symbol are understood to be added in quadrature. In terms of the luminosity, the tagging probabilities and their uncertainties, and after making suitable approximations, we find:
\beq
\frac{\Delta\Vcbsq}{\Vcbsq}\simeq\!\frac{1}{\epsB\sqrt{\sigFac\,\lum\,\epsc\,\epsb}}\oplus\frac{[\flb\!+\!\flc\fcbt(1\!+\!\flc\!+\!\fcbt)]^{\frac{1}{2}}}{\epsB\,\epsb\,\Vcb\sqrt{\sigFac\,\lum\,\epsc}}\oplus\delS\oplus\frac{\flb}{\epsb\,\Vcb^2}\delBa.
\label{relErr}
\eeq
We have combined the terms dependent on $\bg$ and $\delBb\Bb$ in  eq.~(\ref{totErr}) into the term in square brackets, since they are both statistical background contributions, even though they have conceptually-distinct origins. We note that the luminosity moderates only the statistical contributions in eq.~(\ref{relErr}), as expected. The approximate value $\Vcbsq\simeq 0.0018$ is of course already known, although its precise value at $q\simeq m_W$ is unknown, this being the target of the proposed measurement.

The tagging-efficiencies, \epsB, \epsc\ and \epsb, and the mis-tag probabilities, $\fia$, are strongly correlated, lying on efficiency/mis-tag ROC curves, along which we can move in order to minimise the fractional error. In order to get an idea of the performance of the method, we take as an example, our suggested tagging working points from the ATLAS column in table \ref{table:taggingEfficiencies} above, i.e.~$\epsB=\epsb=0.55$ and $\epsc=0.25$. Since we cannot know the future uncertainties on the tagging performances, we start by assuming values close to their current values, i.e.~$\delBa=3\delS=0.1$.
Inserting these values into eq.~(\ref{relErr}), and taking the final \hbox{Run 2} luminosity of $\lum\simeq140\,$fb$^{-1}$, gives a fractional uncertainty on \Vcb\ (i.e.~half that on \Vcbsq):
\bea
\frac{\Delta\Vcb}{\Vcb}
&\simeq&\frac{1}{2}\,(0.086\oplus0.087\oplus0.033\oplus0.061)\cr
&\simeq&0.07,
\label{relErr2}
\eea
where we have written the terms in the same order as they appear in eq.~(\ref{relErr}), to facilitate their interpretation. The first term is the naive signal statistical error; the second term is the background statistical contribution; the third and fourth terms correspond to the systematic uncertainties due to $\delS$ and $\delBa$ respectively. With the Run 2 dataset, the measurement is clearly statistics-limited. We also note that there is a rough balance between the contributions from signal and background, in both the statistical and systematic components. This $7\%$ measurement of \Vcb\ at a single experiment, would be a promising start at this energy scale.

\begin{figure}[t]
\centering
\includegraphics[height=7.8cm]{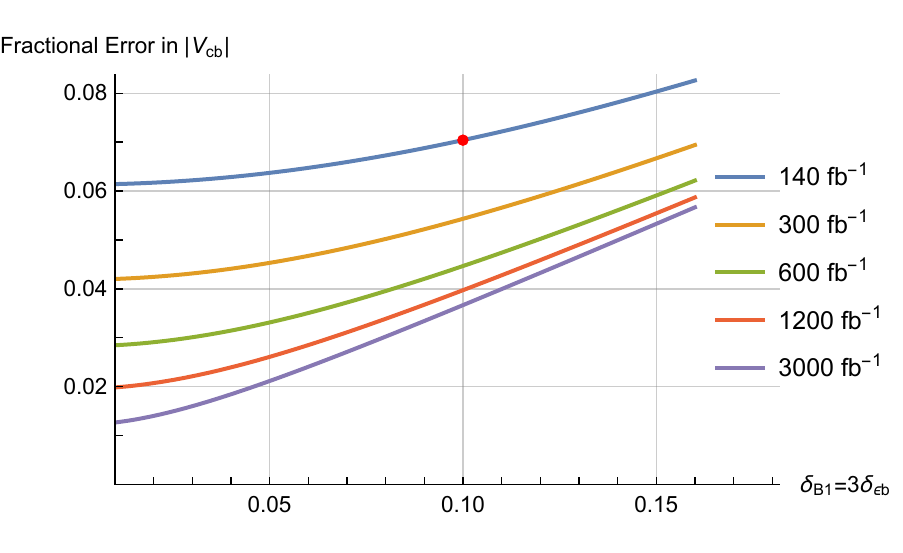}
\caption{Estimated fractional error in $\Vcb$ as a function of systematic tagging efficiency uncertainties, $\delBa=3\delS$ and integrated luminosity. The top curve represents the Run 2 statistics and the red point on it indicates the illustrative values used in eq.~(\ref{relErr2}). The second curve corresponds to luminosity projections for Run 3, while the bottom curve is for the projected integrated luminosity for HL-LHC. We have allowed for a 15\% increase in the $\ttbar$ cross-section in the lower three curves, corresponding to an increase in beam collision energy to 14 TeV.}
\label{sensitivity}
\end{figure}

Since the values chosen for the uncertainties on the tagging performance, $\delS$ and $\delBa$, were based  roughly on their present determinations, we generalise the result of eq.~(\ref{relErr2}) to show in fig.~\ref{sensitivity}, 
how the fractional error on \Vcb\ given by eq.~(\ref{relErr}) depends on these uncertainties as they vary. Purely in order to facilitate their presentation in a two-dimensional figure, we continue with the assumption, based on the currently-obtained values in table \ref{table:taggingEfficiencies}, that $\delBa=3\delS$. Also shown in fig.~\ref{sensitivity} are the results using larger datasets, corresponding to various future LHC luminosity scenarios. The systematics-limited regime is represented by the linear-sloping region towards the bottom-right part of the figure, while the statistics-limited regime lies close to the $y$-axis, where the benefit of more statistics is most marked. Since the example values of eq.~(\ref{relErr2}) indicated by the red point in fig.~\ref{sensitivity} are in the statistics-limited regime, a change in the uncertainties of the tagging probabilities relative to those will make only modest changes to the results. However, at higher luminosities, there is a stronger benefit from improving the uncertainties on the tagging performance. 

Moving to yet tighter tagging regimes could yield further improvements. At this preliminary stage however, it is difficult to explore the benefit of tighter flavour-tagging working points, since this would involve entering territory which has so far not been explored in publications from the LHC experiments. 

\section{Conclusions}
\label{sec:conclusions}
We have proposed a new method to measure the CKM matrix element \Vcb\ at the weak scale at the LHC. We have shown (fig.~\ref{sensitivity}) that taking as a starting point, efficiencies from existing ATLAS and CMS \ttbar\ cross-section analyses, already-achieved experimental tagging performances, and reasonable assumptions about backgrounds, that a measurement of $\Vcb$ at the 7\% level (fractional) per experiment ought to be possible with existing datasets. Making the measurement with future LHC data promises further improvements from both increased statistics and improved tagging performance. E.g., if $\delBa=3\delS$ can be reduced to $\simeq 0.05$, then at the end of Run 3, the uncertainty on \Vcb\ per experiment could be as low as 4.5\%, giving a fractional uncertainty on the average of the two measurements of $\sim3\%$. HL-LHC would then deliver a further reduction in the measurement uncertainty of better than a factor of 2.

Possibilities to extend the method include the use of other channels, such as semi-leptonic \ttbar\ events in which one bachelor $b$ is missed, or perhaps even fully-hadronic \ttbar\ events. Such possibilities bring additional challenges as well as extra statistics, and their study lies outside the scope of this paper.

All previous measurements of $\Vcb$ have been made in $B$ decays at much lower $q^2$. Our proposed method could provide the first measurement of \Vcb\ close to the weak scale and in top decays. If any significant difference is found relative to the existing measurements at low energy scales, this would indicate the presence of new physics.

\vspace{4mm}
{\large
\ni\textbf{Acknowledgements}
}\\
We thank Tim Gershon, Frank Krauss, Michal Kreps, Alex Lenz, Thomas Mannel, Bill Murray and Bill Scott for their helpful comments. 
We also acknowledge support from the Science and Technology Facilities Council (United Kingdom).

\bibliography{vcbPaper}{}
\end{document}